\documentclass{midl} 

\usepackage{mwe} 

\jmlrproceedings{MIDL}{Medical Imaging with Deep Learning}
\jmlrpages{}
\jmlryear{2019}
\jmlrworkshop{MIDL 2019 -- Extended Abstract Track}

\title[Residual Learning and Boundary Weighted Loss for Lesion Segmentation]{Significance of Residual Learning and Boundary Weighted Loss in Ischaemic Stroke Lesion Segmentation}

\midlauthor{\Name{Ronnie Rajan\midljointauthortext{Contributed equally}\nametag{$^{1}$}} 
\Email{dr.ronnierajan@iitkgp.ac.in}\\
\addr $^{1}$ School of Medical Science and Technology, Indian Institute of Technology Kharagpur
\AND
\Name{Rachana Sathish\midlotherjointauthor\nametag{$^{2}$}} 
\Email{rachana.sathish@iitkgp.ac.in}\\
\Name{Debdoot Sheet\nametag{$^{2}$}} \Email{debdoot@ee.iitkgp.ac.in}\\
\addr $^{2}$ Department of Electrical Engineering, Indian Institute of Technology Kharagpur
}

\begin{document}

\maketitle

\begin{abstract}
Radiologists use various imaging modalities to aid in different tasks like diagnosis of disease, lesion visualization, surgical planning and prognostic evaluation. Most of these tasks rely on the the accurate delineation of the anatomical morphology of the organ, lesion or tumor. Deep learning frameworks can be designed to facilitate automated delineation of the region of interest in such cases with high accuracy. Performance of such automated frameworks for medical image segmentation can be improved with efficient integration of information from multiple modalities aided by suitable learning strategies. In this direction, we show the effectiveness of residual network trained adversarially in addition to a boundary weighted loss. The proposed methodology is experimentally verified on the SPES-ISLES 2015 dataset for ischaemic stroke segmentation with an average Dice coefficient of $0.881$ for penumbra and $0.877$ for core. It was observed that addition of residual connections and boundary weighted loss improved the performance significantly.
\end{abstract}


\section{Introduction}
Medical imaging tools play a vital role in aiding the physician in various tasks like estimating the anatomical morphology of organ for surgical planning, lesion visualization to evaluate damage and assessment of tumor size and spread for excision or prognostic ranking, etc. Thus precise demarcation of a lesion or organ plays a crucial role in planning and deciding life-saving therapy. One such example is the evaluation of acute ischaemic stroke and delineation of the extent of necrotic \textit{core} in the centre of the lesion and salvageable \textit{penumbra} around it\cite{dirnagl1999pathobiology}. Decision on thrombolytic therapy that can reverse the damage in the penumbra and thus alleviate associated symptoms, depends on the accurate estimation of extent of these lesions\cite{atlantis2004association}. To this regard, it is essential that all frameworks for automated semantic segmentation in medical imaging must include strategies that penalizes the misclassification of the boundary pixels heavily.

\section{Methodology and Experiments}
The proposed method uses SUMNet \cite{nandamuri2019sumnet} as the base model for segmentation with modifications. We do not use ImageNet pre-trained weights, instead train the model from scratch with the addition of batch-normalization on the ISLES dataset\footnote{http://www.isles-challenge.org/ISLES2015/} using three MRI sequences available in the SPES subset of the dataset, \textit{viz.} TMax, TTP and DWI. These three sequences are concatenated into a tensor and given as the input to the network. We adopt an adversarial training strategy similar to our recent work \cite{sathish2019adversarially} where we employed three relativistic discriminators operating on the segmented core, segmented penumbra and the pair respectively. In this paper we present an improvement in performance caused by the addition of residual connections \cite{he2016deep} in the network along with a boundary weighted loss.
In the modified architecture, we add residual connections in the VGG11 \cite{simonyan2014very} like encoder after each convolutional block as shown in Figure. The network is trained using multiple losses which include Cross-Entropy (CE) loss, Adversarial loss (Adv. loss) from the discriminators, Lovasz-Softmax (LS) loss \cite{berman2018lovasz} and weighted Negative Log-likelihood based boundary loss (BD) for the boundary pixels. A 3-pixel thick boundary is extracted from the ground truth annotation by subtracting the eroded ground truth from the dilated one using a structuring element of size $3\times3$. The pixels in this boundary are then weighed by a factor of $10$. 

\begin{figure}[h]
    \centering
    \includegraphics[width=\textwidth]{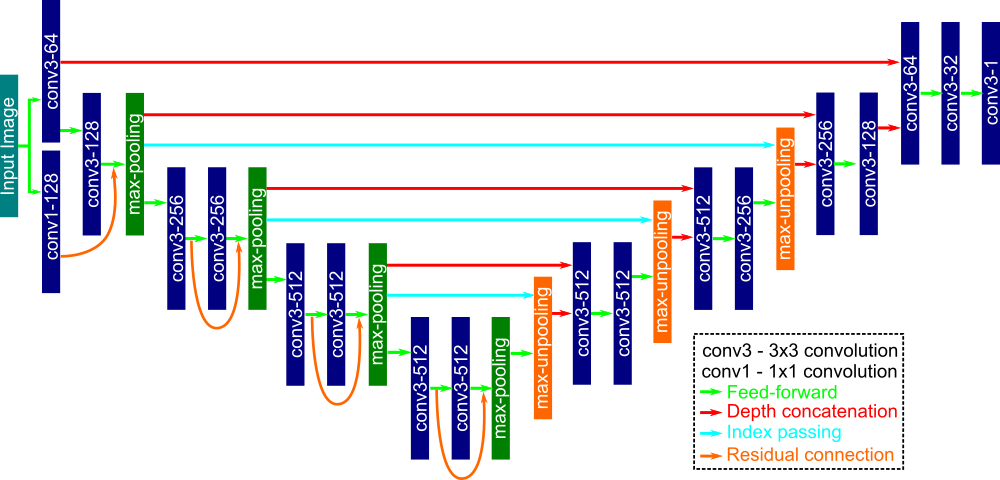}
    \caption{Architecture of the proposed network}
    \label{fig:my_label}
\end{figure}

Performance of the proposed method is compared with the following baselines:\\
\noindent \textbf{BL1:} SUMNet trained using only CE loss\\
\noindent \textbf{BL2:} SUMNet with CE, LS and BD losses\\
\noindent \textbf{BL3:} SUMNet trained adversarially along with CE loss\\
\noindent \textbf{BL4:} SUMNet trained adversarially along with CE, LS and BD losses\\
\noindent \textbf{BL5:} Residual-SUMNet trained using only CE loss\\
\noindent \textbf{BL6:} Residual-SUMNet trained with CE, LS and BD losses\\
\noindent \textbf{BL7:} Residual-SUMNet trained adversarially along with CE loss\\
\noindent \textbf{Proposed:} Residual-SUMNet trained adversarially along with CE, LS and BD losses

\section{Results and Discussion}
Three-fold cross validation is performed to evaluate various baselines and the performance, as measured by average Dice-coefficient across the folds, is presented in Tab.\ref{tab:expt}. 

\begin{table}[h]
\centering
\caption{Performance Evaluation of Proposed Method in terms of Dice-coefficient}
\label{tab:expt}
\begin{tabular}{|c|c|c|c|c|c|c|c|c|}
\hline
         &  BL1  &  BL2  &  BL3  &  BL4  &  BL5  &  BL6   &  BL7 & \textbf{Proposed}   \\ \hline
Penumbra & 0.835 & 0.838 & 0.803 & 0.841 & 0.845 & 0.844 & 0.852 & \textbf{0.881} \\ \hline
Core     & 0.792 & 0.802 & 0.730 & 0.813 & 0.867 & 0.874 & 0.865 & \textbf{0.877} \\ \hline
\end{tabular}
\end{table}

The addition of residual connections (BL5) to SUMNet (BL1) increases the Dice coefficient from $0.835$ to $0.845$ for penumbra and $0.792$ to $0.867$ for core. Similar trend is seen in the baselines with adversarial training (BL3 and BL7), with or without additional losses (BL2-BL6 and BL4-Proposed), as shown in Tab.~\ref{tab:expt}. The significant improvement in segmentation of the core as compared to penumbra can be attributed to the residual connections. The core being smaller, much of the information is lost along the depth of the network in the absence of residual connections.

Further improvement in performance is observed when LS and BD losses are added in conjunction with CE loss. This increase in performance is noted between BL1 and BL2 with a Dice coefficient of $0.835$ to $0.838$ for penumbra and $0.792$ to $0.802$ for core. Tab.~\ref{tab:expt} shows the corresponding improvement in performance in BL4, BL6 and Proposed method as compared to BL3, BL5 and BL7 respectively. 

\begin{figure}[h]
    \centering
    \subfigure[TMax]{\includegraphics[width=0.223\textwidth
    ]{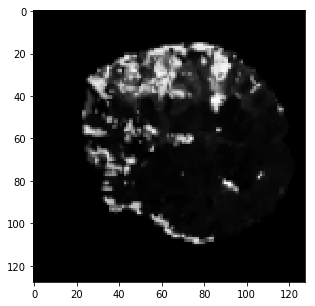}}
    \subfigure[TTP]{\includegraphics[width=0.223\textwidth]{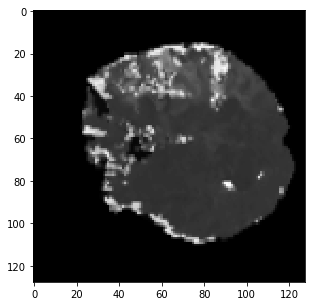}}
    \subfigure[DWI]{\includegraphics[width=0.223\textwidth]{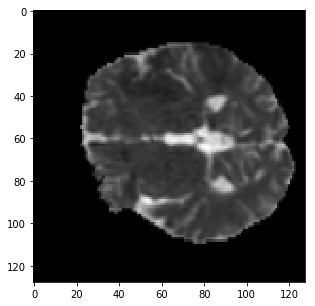}}
    \subfigure[GT]{\includegraphics[width=0.223\textwidth]{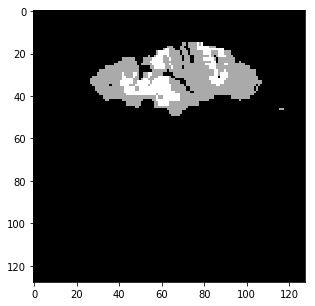}}
    
    \subfigure[BL3]{\includegraphics[width=0.223\textwidth]{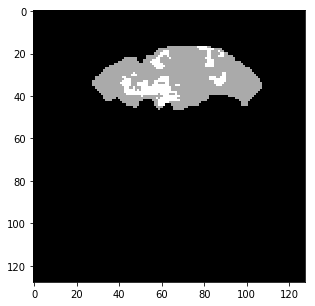}}
    \subfigure[BL4]{\includegraphics[width=0.223\textwidth]{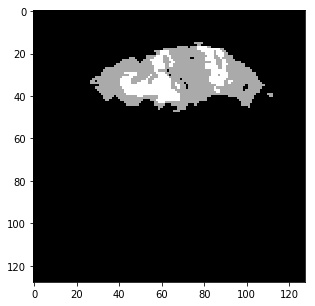}}
    \subfigure[BL7]{\includegraphics[width=0.223\textwidth]{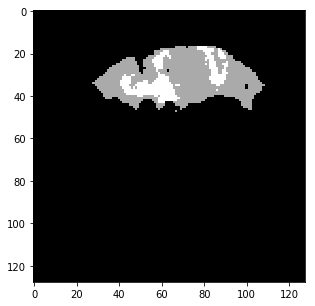}}
    \subfigure[Proposed]{\includegraphics[width=0.223\textwidth]{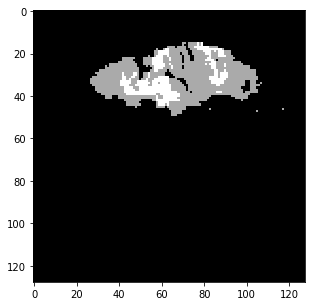}}
    
    \caption{Figure shows sample inputs, ground truth annotation and the segmented results (e)-(h) for different baselines (BL) and the proposed  method.}
    \label{fig:results}
\end{figure}

The qualitative results are shown in Fig.~\ref{fig:results}. Addition of LS and BD losses while training the network significantly improves the boundary delineation in the results. This improvement is more drastic when the network is adversarially trained as seen in the results of the proposed method as compared to BL7.

It can be seen from Tab.~\ref{tab:expt}, the performance increases significantly with the addition of residual connection and the boundary loss. This is also evident from the qualitative results. The residual connections in the network helps in propagation of multi-sequence information through the network. Also, the weighted boundary loss improves the segmentation along the boundary of the lesion. 

\section{Conclusion}
Residual connections in deep neural networks significantly improve the performance of very deep neural networks in the task of classification. In this work, we evaluate it's effectiveness for semantic segmentation in medical images. With the addition of a boundary weighted loss, the boundaries of the different regions of interests are more accurately predicted by the proposed method which adds value from a clinical perspective.

\bibliography{midl}
\end{document}